\documentclass{article}

\usepackage{amsthm,amsmath, amsfonts, amssymb}
\usepackage{graphicx, color, tikz}
\usepackage{cite}
\usepackage{times}
\usepackage{fullpage}
\usepackage{mathptmx}
\usepackage{latexsym}

\usetikzlibrary{arrows}

\newtheorem{theorem}{Theorem}
\newtheorem{lemma}[theorem]{Lemma}

\newtheorem{fact}[theorem]{Fact}
\newtheorem{myclaim}[theorem]{Claim}

\newcommand{\ps}{{\sc PS~}}
\newcommand{\gps}{{\sc GPS~}}
\newcommand{\gpsX}{{\sc GPS}}
\newcommand{\gpsalgo}{{\sc GPSAlgo~}}

\newcommand {\bbZ}    {\mathbb{Z}}

\title{Precedence-constrained Scheduling of Malleable Jobs with Preemption\thanks{Part of this work was done when both authors were at Microsoft Research, Redmond, WA.}}
\author{Konstantin Makarychev\\Microsoft Research\\Redmond, WA\\{\tt komakary@microsoft.com} \and
Debmalya Panigrahi\thanks{Supported in part by a Duke University startup grant.}\\ Duke University\\ Durham, NC\\{\tt debmalya@cs.duke.edu}}

\date{}

\begin{document}

\maketitle

\begin{abstract}
Scheduling jobs with precedence constraints on a set of identical machines
to minimize the total processing time (makespan) is a fundamental problem 
in combinatorial optimization. In practical settings such as cloud
computing, jobs are often {\em malleable}, i.e., can be processed on multiple machines 
simultaneously. The instantaneous processing rate of a job is a 
non-decreasing function of the number of machines assigned to it
(we call it the processing function). 
Previous research has focused on practically relevant
concave processing functions,
which obey the law of diminishing utility and generalize the classical 
(non-malleable) problem.
Our main result is a $(2+\epsilon)$-approximation algorithm 
for concave processing functions (for any $\epsilon > 0$),
which 
is the best possible under complexity theoretic assumptions.
The approximation ratio improves to $(1 + \epsilon)$ for the interesting
and practically relevant special case of power functions, i.e., 
$p_j(z) = c_j \cdot z^{\gamma}$. 
\end{abstract}

\section{Introduction}
\label{sec:intro}

In the {\em precedence-constrained scheduling} problem (we call it 
the \ps problem), the goal is to schedule a set of jobs with precedence 
constraints on a set of identical machines so as to minimize the overall
time for processing them (called the {\em makespan}). One of the first
results in approximation algorithms was a 2-approximation for this problem
due to Graham in 1966~\cite{Graham66}. On the negative side, this problem 
was shown to be NP-hard to approximate to a ratio 
better than 4/3 by Lenstra and Rinnooy Kan in 1978~\cite{LenstraK78}. 
In spite of substantial effort, the gap between these two bounds 
remained open for three decades. Recently, 
Svensson~\cite{Svensson11} has provided strong evidence that 
improving Graham's result might in fact be impossible by showing that it is tight 
under certain complexity theoretic assumptions. 
%We discuss this problem and 
%its variants in more detail in related work.

A natural generalization of the \ps problem considered in the literature
is that of {\em malleable} jobs, i.e., jobs that can be processed 
simultaneously on multiple machines. This is particularly relevant
in practice for domains such as cloud computing, 
% typically the case in many 
%applications. Consider, for example, a distributed computing environment
%such as a cloud computing center. The scheduler is presented with a 
%set of precedence-constrained jobs, and must use its fixed processing 
%resources to serve these jobs. Similar situations arise in 
operating systems, high performance computing, project management, etc.
where a fixed set of resources must be distributed among
precedence-constrained tasks to complete them by the earliest possible 
time. At any given time, the processing rate of a job is a function 
of the number of machines assigned to it (we call it the 
{\em processing function}). The goal is to produce a schedule of 
minimum makespan. 
%We call this the 
%{\em generalized precedence-constrained scheduling} problem. 

Formally, the input comprises a {\em directed acyclic graph} (DAG) $G = (J, E)$,
%%\kostya: Replaced $A$ with $E$ to avoid confusion with the schedule $A$.
where each vertex $j\in J$ represents a job $j$ and has a given size 
$s_j > 0$. The arcs in $E$ represent the {\em precedence constraints}
on the jobs, i.e., if $(j_1, j_2)\in E$, then job $j_1$ has to be 
completed before job $j_2$ can be processed. Let $m$ denote the number of
identical machines on which these jobs have to be scheduled.
We are also given processing functions for the jobs
$p_j: \{0, 1, 2, \ldots, m\} \rightarrow \mathbb{R}^+_0$
that map the number of machines assigned to the rate at 
which the job is processed. (Clearly, $p_j(0) = 0$ for all 
processing functions.) 
%We call this the {\em processing function} for job $j$.

The output of the algorithm is a schedule $A$,
which is represented by a continuum of functions $A_t(j)$
over time $t > 0$. $A_t(j)$ represents the number
of machines allocated to job $j$ at time $t$. The schedule must satisfy:
\begin{itemize}
	\item {\em Capacity constraints}: For any time $t\in (0, \infty)$, 
	the number of allocated machines 
	at time $t$ is at most $m$, i.e.,	$\sum_{j\in J} A_t(j) \leq m$.
	\item {\em Precedence constraints}:
	For any arc $(j_1, j_2)\in E$ and any time $t\in (0, \infty)$, 
	if $A_{t}(j_2) > 0$, then the job $j_1$ must be finished by time $t$, 
	i.e. $\int_{0}^{t} p_{j_1}(A_{t'}(j_1)) dt' \geq s_{j_1}$.
	Let the set of jobs that can be processed at a given time (i.e., all
	their predecessors in $G$ have been completely processed) be called the 
	set of {\em available} jobs. Then, the precedence constraints enforce
	that the schedule picks a subset of available jobs to process at 
	any given time.
\end{itemize}	
The {\em makespan} (or length) of the schedule is defined as the
time when all jobs finish processing, i.e., 
$\ell(A) = \sup \{t: \sum_{j\in J} A_t(j) > 0\}$.
The objective of the algorithm is to minimize the makespan of the
schedule. We call this the 
{\em generalized precedence-constrained scheduling}
%problem}, or the 
or \gps problem.

\medskip
\noindent
{\bf Preemption.}
It is important to note that we allow {\em preemption}, i.e., at any point of 
time, the remaining volume of an available job can be scheduled on any number 
of machines independent of the history of where it was processed earlier. 
Therefore, our schedule is defined simply by the number of machines allocated
to a job at any given time, and not by the identities of the machines themselves.
This is a departure from the bulk of the existing literature in 
precedence-constrained scheduling with malleable jobs, where preemption is 
typically disallowed. However, our motivation for allowing preemption comes from 
the fact that it is allowed in many application domains (such as scheduling in 
cloud computing) and has been widely considered in the broader scheduling literature.

\medskip
\noindent
{\bf Processing Functions.} 
Following \cite{PrasannaM91, JansenZ12},
we consider processing functions that are 
(1) {\em non-decreasing} (assigning more 
machines does not decrease the rate of processing) and 
(2) {\em concave}\footnote{We note that
it is optimal to process an arbitrary available job on all machines 
simultaneously if all processing functions are {\em convex}.} (the 
processing rate obeys the law of diminishing marginal utility because 
of greater overhead in coordination, communication costs, etc. between
the machines processing a job).\footnote{Other classes of processing 
functions have also been considered in the literature (see related work), 
but monotonicity and concavity are two basic qualitative features of
processing functions in most applications.} 
Note that concave processing functions generalize the classical \ps problem
%
%. In the \ps problem, we are also given a DAG on jobs (of length
%1, wlog) and a set of $m$ identical machines, but 
%the processing functions are of the following form:
($p_j(z) = 1$ iff $z \geq 1$ for all jobs $j$).
%As in our problem, we can scale the number of machines to 1
%and use Lemma~\ref{lma:continuous} to rewrite the processing 
%functions as
%$p(x) = m x$ for $x < 1/m$ and $p(x) = 1$ for $x \geq 1/m$.
%Note that this function is concave; hence, our problem generalizes
%the \ps problem.
%We note that the \ps problem has a greedy 2-approximation algorithm
%that has not been improved upon for almost fifty years~\cite{Graham66}
%and may be impossible to improve based on recent evidence~\cite{Svensson11}.

%\iffalse
%\footnote{Sometimes,
%the classical problem is defined on a discrete time interval, whereas 
%Lemma~\ref{lma:continuous} crucially uses a continuous time interval,
%i.e., the ability of machines to switch jobs at any time rather than at
%discrete timepoints. If we are limited to a discrete time interval, then
%we lose a factor of 2 in simulating a continuous schedule by a discrete
%one.\debmalya{Can we do a randomized simulation that loses only $o(1)$?}}\fi

%%\begin{enumerate}
%%\item 
% We will assume, by scaling, that the total number of machines
%	$m = 1$, and allow fractional assignment of machines to jobs. 
%	Accordingly, the processing function is now defined as
%	$p: \{0, 1/m, 2/m, \ldots, 1\} \rightarrow \mathbb{R}^+_0$. 

\medskip
\noindent
{\bf Integrality of machines.}
The existing literature is divided between allowing fractional 
allocation of machines to jobs
(e.g. \cite{PrasannaM91, PrasannaM94}) or
enforcing integrality of machine allocations 
(e.g., \cite{JansenZ12, JansenZ06, LepereTW02}).
Accordingly, the processing functions are defined on the entire
interval $[0, m]$ or on the discrete values $\{0, 1, \ldots, m\}$.
Since we allow job preemption, a fractional 
assignment of machines to jobs can be realized by a round robin 
schedule even if there is no inherent support in the application
for jobs to share a machine. 
%Moreover, many of the applications that motivated
%our work are inherently equipped with resource sharing mechanisms,
%e.g. threads to share a processing unit among multiple jobs.
Therefore, if the processing function is defined only on an
integral domain, we extend it to the continuous domain by linear
interpolation between adjacent points. In the rest of the paper, 
we will assume that the 
processing functions $p_j(.)$ are defined on the continuous domain
$[0, m]$ and fractional schedules are valid.
%
%	We will extend the processing function to a continuous function
%	by linearly extending the function between any two discrete points
%	to obtain $p: [0, m] \rightarrow \mathbb{R}^+_0$.
%	The next lemma ensures that this extension is wlog.
%	\begin{lemma}[see Section~\ref{lem:one} in the appendix]
%		\label{lma:continuous}
%		Any schedule that uses the continuous version of the 
%		processing functions can be simulated by a schedule that
%		uses the discrete version without any increase in the 
%		total length of the schedule.
%	\end{lemma}
	
%%	\item 
%  We will also assume that the size of every job is 1. 
%	Note that a job $j$ of size $s_j$ is equivalent to a directed path
%	of $s_j$ jobs of size 1 each in graph $G$
%	(after appropriate scaling to ensure integrality of $s_j$). Hence, this 
%	assumption is w.l.o.g.
%%\end{enumerate}	

\medskip
\noindent
{\bf Our Results.}
%
%This paper will focus on {\em concave} processing functions, which 
%represent the principle of diminishing utility. However, for completeness,
%let us first consider {\em convex} processing functions. The next property is 
%straightforward.
%\begin{fact}
%	If the processing functions are convex, then there exists an optimal
%	solution where at any point of time, only one job is processed, i.e.,
%	at time $t$, $A_t(j) = m$ for some job $j$ and 0 for all other jobs.
%\end{fact}
%Using the above fact, we can immediately
%claim that the following algorithm is optimal for convex processing 
%functions:
%At any time, choose an arbitrary available job and assign all
%$m$ machines to this chosen job. Once this job completes, 
%update the set of available jobs and repeat.
%
%
Our main result is a $(2+\epsilon)$-approximation algorithm for the \gps
problem for concave processing functions. Note that this matches the best 
known bounds for the \ps problem.

\begin{theorem}
\label{thm:concave}
	For any $\epsilon > 0$, there is a deterministic algorithm \gpsalgo 
	for the \gps problem that has an approximation factor of $(2+\epsilon)$
	for concave processing functions. 
\end{theorem}
We note that if preemption is disallowed, then the best approximation 
ratio known is 3.29 due to Jansen and Zhang~\cite{JansenZ12}.

In practice, a particularly relevant set of processing functions are {\em power functions}
(for examples of their practical importance, see, e.g., \cite{PrasannaM91}), 
i.e., $p_j(z) = c_j \cdot z^{\gamma}$ for $c_j > 0$. We show that
our algorithm is in fact {\em optimal} for this special case. (Note that 
(1) power functions do not generalize the \ps problem and
(2) while the multiplier $c_j$ can depend on the job, the exponent $\gamma$ 
in the power functions has to be universal for our analysis.)

\begin{theorem}
\label{thm:power}
	For any $\epsilon > 0$, \gpsalgo has 
	an approximation factor of $(1+\epsilon)$ if the processing functions are
	power functions.
\end{theorem}

\medskip
\noindent
{\bf Our Techniques.}
It would be natural to try to extend the greedy approach of 
Graham's algorithm for the \ps problem to our problem. The basic
scheduling rule of Graham's algorithm is the following: if there
is an idle machine and an available job that is currently not 
being processed, then schedule this job on the machine. 
%
%
%Our algorithm is quite different from Graham's greedy
%algorithm~\cite{Graham66} for the \ps problem. 
Note that this algorithm is {\em online} in the sense that
it can operate on an instance where a job is revealed only 
after it becomes available. We categorically refute the 
possibility of extending this greedy approach to the 
\gps problem by giving a polynomial lower bound
on the approximation factor obtained by {\em any} online 
algorithm for the \gps problem. 

\begin{theorem}
\label{thm:online}
	No online algorithm for the \gps problem can have a sub-polynomial 
	competitive ratio, even if all the processing functions are a 
	fixed power function.
\end{theorem}

Instead, we employ an LP rounding approach (following the work of 
Chudak and Shmoys~\cite{ChudakS99} and Skutella~\cite{Skutella97}). 
In designing the LP relaxation, 
we introduce a variable $x_{ja}$ denoting the duration for which job $j$ 
is processed simultaneously by $a$ machines. Then, the processing time 
for job $j$ is $X_j = \sum_a x_{ja}$. The goal is then to minimize
the makespan $T$ subject to the following constraints: 
(1) the total processing time of jobs on any chain (maximal directed path 
in the precedence graph) $\sum_{j\in C} X_j$ is a lower bound on the 
makespan $T$;
(2) the total number of machine-hours for all jobs $\sum_j \sum_a a \cdot x_{ja}$
is a lower bound on $mT$; and
(3) the total processing volume for any single job $\sum_a p_j(a)\cdot x_{ja}$
is at least its size $s_j$. 
%
%
%We show that this LP can be solved using 
%a $(2+\epsilon)$ approximate dual separation oracle. The algorithm
%solves the LP and obtains a schedule which satisfy non-interference 
%constraints but may violate precedence constraints. 
%We note that this LP formulation is somewhat similar to one used by 
%Chudak and Shmoys~\cite{ChudakS99} for a different generalization of 
%the \ps problem to machines with non-identical speeds. However, this 
%similarity is somewhat superficial. In our problem, the optimal LP 
%solution does not provide a schedule (even a fractional one), and the 
%main challenge is to show that any set of variables $x_{ja}$ that satisfy
%the LP constraints can be converted to a schedule for processing the 
%jobs losing only a factor of 2 in the makespan objective. On the other
%hand, the LP solution in \cite{ChudakS99} provide a feasible schedule,
%but it is a fractional one. Therefore, their main challenge was to 
%round these fractional variables to integer values, for which they
%gave an algorithm with a logarithmic loss in the makespan objective.

First, we solve our LP to obtain optimal values of $x_{ja}$'s. Next, 
we structure this solution by showing that $x_{ja} = 0$ for all 
{\em except one} value of $a$ for each job $j$ in an optimal solution
w.l.o.g. Let us call this value $b^*_j$. We now create a feasible 
schedule by using the following simple rule: 
at any given time, we distribute
the available jobs among all the $m$ machines in proportion to their 
values of $b^*_j$. The key property that we use in the analysis is the 
following: (1) if there are too few available jobs (quantified by 
the sum of $b^*_j$'s of available jobs being less than $m$), then the
non-decreasing property of the processing function ensures that for 
every chain, at least one job is being processed faster than in the 
LP solution, and (2) if there are too many available jobs (quantified 
by the sum of $b^*_j$'s of available jobs being greater than $m$), 
then the concavity of the processing function ensures that 
remaining overall ratio of the job volumes and machine-hours is
decreasing at a faster rate than in the optimal LP solution.
These two observations lead to the conclusion that the makespan 
of the schedule is at most twice the LP objective.

For the class of power functions, i.e.,
$p_j(z) = c_j\cdot z^{\gamma}$ (we only consider $\gamma \in [0,1]$
since otherwise, the function is convex for which we have already 
shown that there is a simple optimal algorithm) our LP is exact
(up to a factor of $(1+\epsilon)$ for any $\epsilon > 0$). 
The main insight is that the special structure of power functions
allows us to employ simple linear algebraic inequalities to design
a function that trades off the two cases above. More precisely, 
we show that the gains/losses made by the algorithm 
over the optimal LP solution for the processing time of chains
are exactly compensated by the losses/gains made by it 
over the LP solution for the overall number of machine-hours
in the two situations described above.

%Furthermore, we show that we can solve the LP up to an arbitrarily
%small multiplicative error $1+\epsilon$ ($\epsilon$ can be made exponentially small; it
%is not possible to solve the LP exactly, since the optimal solution may be irrational).

%Finally, we also show that Theorem~\ref{thm:concave} represents a limit
%on our approach in that there exists an instance of the \gps problem where
%a gap of 2 can be established between the optimal LP objective and the 
%minimum makespan. We call this gap the {\em precedence gap} of our LP 
%formulation (rather than the more commonly used term {\em integrality gap} 
%since integrality is not what separates the LP solution from the schedule).
%
%\begin{theorem}
%\label{thm:gap}
%	There is a precedence gap of 2 for the LP formulation of the \gps
%	problem that we use in Algorithm \gpsalgo.
%\end{theorem}

\medskip
\noindent
{\bf Related Work.}
The precedence-constrained scheduling problem with malleable jobs has a 
long history in approximation algorithms. Du and Leung~\cite{DuL89}
showed that the problem is NP-hard even for a small number of machines
and gave optimal algorithms if the precedence graph has special structure.
Turek~{\em et al}~\cite{TurekWY92} 
considered the problem of scheduling malleable tasks 
in the absence of precedence constraints and obtained approximation 
algorithms for both the preemptive and non-preemptive situations.
%for more general processing functions.
In the presence of precedence constraints, 
several families of processing functions have been considered.
Our model was originally suggested by 
Prasanna and Musicus~\cite{PrasannaM91, PrasannaM94, PrasannaM96} and 
subsequently used by Jansen and Zhang~\cite{JansenZ12}, who 
obtained as approximation factor of 3.29 for 
the non-preemptive version of our problem. In some papers,
the concavity requirement is replaced 
by a weaker constraint that the size of the 
jobs increases as more machines are assigned to 
it~\cite{JansenZ06,LepereTW02}. For this model, the best 
known approximation factor is 4.73~\cite{JansenZ06}. A third
(more general) model that has been considered is that of arbitrary speed up 
curves. Here, the processing rate not only depends on the number
of assigned machines and the job being processed, but also 
on the stage of processing of a job~\cite{Edmonds00, EdmondsCBD03}.
Most of the literature in this model is geared toward minimizing
the flow-time (rather than the makespan) (see 
e.g.~\cite{ChanEP11, EdmondsP12, FoxIM13}), 
including in the presence of precedence constraints~\cite{RobertS08}.
%Moreover, their setting was non-clairvoyant, but as we show in 
%this paper, even an online model is too restrictive for our problem. 
For a detailed survey on scheduling parallelizable jobs, the 
reader is referred to \cite{DutotMT04}.

%As mentioned earlier, the \ps problem was one of the first optimization
%problems for which an approximation algorithm with a provable guarantee
%was shown. After Graham~\cite{Graham66} showed an approximation factor 
%of $2-1/m$, several algorithms obtaining better bounds for small values of 
%$m$ have been proposed. 
%Lam and Sethi~\cite{LamS77}\footnote{An error in this analysis was 
%corrected by Braschi and Trystram~\cite{BraschiT94}.} showed that an
%algorithm by Coffman and Graham~\cite{CoffmanG72}  has and approximation
%factor of $2-\frac{2}{m}$, which was improved recently by Gangal and 
%Ranade~\cite{GangalR08} to $2 - \frac{7}{3m+1}$. 
%On the negative side, the results
%of Lenstra and Rinnooy Kan~\cite{LenstraK78} and Svensson~\cite{Svensson11}
%establish inapproximability for factors better than 4/3 and 2 respectively
%based on corresponding complexity theoretic assumptions. 

Since the work of Graham, both upper bounds (particularly, the trailing $o(1)$
factor in the approximation ratio) (see, e.g.,~\cite{LamS77,GangalR08}) 
and lower bounds~\cite{LenstraK78, Svensson11} 
for the \ps problem have been extensively studied. Moreover, 
multiple variants of this problem have been considered. 
This includes optimizing for other 
metrics such as completion time (see, e.g., \cite{BansalK09} and 
references contained therein), handling machines with non-identical
speeds~\cite{ChudakS99, ChekuriB01}, dealing with online 
input~(e.g., \cite{HuoL05}), etc.
For a more detailed history of precedence constrained scheduling,  
the reader is referred to the surveys of 
Graham~{\em et al}~\cite{GrahamLLK79}
and Chen~{\em et al}~\cite{ChenPW98}. 

%%\debmalya{Something about configuration LPs.}

%\medskip
%\noindent
%{\bf Roadmap.} In Section~\ref{sec:concave}, we present the details of 
%\gpsalgo and its analysis for general concave functions, thereby proving
%Theorem~\ref{thm:concave}. We refine this analysis for power functions 
%in Section~\ref{sec:power}, and show Theorem~\ref{thm:power}. We present
%our online lower bound (Theorem~\ref{thm:online}) and the precedence gap
%example (Theorem~\ref{thm:gap}) in Sections~\ref{sec:online} and 
%\ref{sec:gap} in the appendix, respectively.

%\pagebreak

\section{Linear Program}
In this section, we give a linear programming relaxation for the problem. In the discrete case,
when the optimal solution allocates only an integral number of machines to each jobs, we let $A=\{1,\dots, m\}$.
In the continuous case, when the number of machines can be any real number from $[0,m]$, 
We pick an $\varepsilon>0$, and let $A=\{(1-\varepsilon)^k\in [0,m]: k\in \bbZ\}$. Now for every 
job $j\in J$ and every value $a\in A$, we introduce a variable $x_{ja}$. In the intended solution corresponding to
the optimal solution of \gpsX, $x_{ja}$ is equal to the amount of time at which the number of machines used by the
job $j$ is between $(1-\varepsilon) a$ and $a$ (in the discrete case, $\varepsilon = 0$). We let $T$ be the makespan of the schedule. Our goal is to
minimize $T$. We write two constraints on $T$ that are satisfied in the optimal solution.

To write the first constraint, we consider an arbitrary chain of jobs $C$. All jobs $j\in C$ must be processed
sequentially one after another. It takes at least $\sum_{a\in A} x_{ja}$ amount of time to finish 
job $j$. Thus, for every chain $C$,
\begin{equation}\label{eq:lb1}
T \geq \sum_{j\in C}\sum_{a\in A} x_{ja}.
\end{equation}

To write the second constraint, we count the number of machine hours used by the optimal solution. On one hand, 
every job $j$ uses at least 
$\sum_{a\in A} (1-\varepsilon) a x_{ja}$
machine hours. So the total number of machine hours is lower bounded by 
$\sum_{j\in J}\sum_{a\in A} (1-\varepsilon) a x_{ja}$.
On the other hand, the number of machine hours is upper bounded by $m T$. So 
we have
\begin{equation}
m T\geq \sum_{j\in J}\sum_{a\in A} (1-\varepsilon) a x_{ja}.
\end{equation}
To simplify notation, we let $\tilde T = T/(1-\varepsilon)$. 
We finally add constraint (\ref{LP3}) saying that every job $j$ is completed in the optimal solution.
%%for all~$j\in J$,
%%$$\sum_{a\in A} x_{ja} \,p_j(a) \geq s_j.$$
We obtain the following LP relaxation.

%\OpenFrame
\begin{align}
\nonumber\mathbf{minimize}\;&\;\tilde T\quad \mathbf{subject~to}& \\
%\intertext{\textbf{subject to}}
\sum_{j\in C}\sum_{a\in A} x_{ja}&\leq \tilde T &\text{for every chain } C\label{LP1}\\
\sum_{j\in J}\sum_{a\in A}  a\, x_{ja}&\leq \tilde T m&\label{LP2}\\
\sum_{a\in A} x_{ja} \,p_j(a) &\geq s_j&\text{for every job } j\in J\label{LP3}\\
x_{ja}&\geq 0&\text{for all } j\in J,\,a\in A\label{LP4}
\end{align}
%\CloseFrame

Since the LP solution corresponding to the optimal solution satisfies all the constraints of
the linear program, we have $LP\leq OPT$, where $LP$ is the cost of the optimal LP solution, and 
$OPT$ is the cost of the optimal combinatorial solution.

This linear program has infinitely many variables and exponentially many constraints, but using
standard methods we can solve this linear program up to any precision $(1+\varepsilon')$
in polynomial-time. %For details, the reader is referred to the full version of the paper~\cite{full}.
We give the details in Appendix~\ref{ap:solveLP}.

\section{Simplified LP Solution}

It turns out, that every solution to the LP (\ref{LP1}--\ref{LP4}) can be converted to another
simpler solution in which for every job $j$ one and only one $x_{ja}$ is nonzero. Suppose
$x^*_{ja}$ is the optimal solution to the LP (\ref{LP1}--\ref{LP4}), define $y^*_j$'s and $b^*_j$'s
as follows:
\begin{equation}
y^*_j = \sum_{a\in A} x^*_{ja};\;\;\;\;
b^*_j = \frac{1}{y^*_j}\sum_{a\in A} x^*_{ja} a.\label{eq:def-yb}
\end{equation}

\begin{myclaim}\label{claim:yb}
Variables $y^*_j$ and $b^*_j$ satisfy the following constraints %(similar to (\ref{LP1}--\ref{LP4})):
(similar to (\ref{LP1}--\ref{LP3})):
\begin{align*}
\sum_{j\in C} y^*_j&\leq \tilde T &\text{for every chain } C\tag{$\ref{LP1}'$}\\
\sum_{j\in J} \, b^*_j y^*_j&\leq \tilde T m&\tag{$\ref{LP2}'$}\\
y^*_j\,p_j(b^*_j) &\geq s_j&\text{for every job } j\in J\tag{$\ref{LP3}'$}%\\
%%y^*_j &\geq 0&\text{for every job } j\in J\tag{$\ref{LP4}'$}
\end{align*}
\end{myclaim}
\begin{proof}
For every chain $C$, we have
$\sum_{j\in C} y^*_j = \sum_{j\in C} \sum_{a\in A} x^*_{ja} \leq \tilde T$.
Then,
$$\sum_{j\in J} b^*_j y^*_j = 
\sum_{j\in J} y^*_j \cdot \frac{1}{y^*_j}\sum_{a\in A} x^*_{ja}= 
\sum_{j\in J}\sum_{a\in A} x^*_{ja}\leq \tilde T m.$$
Finally, for every $j$, we have
$y^*_j\,p_j(b^*_j) = y^*_j\,p_j\Big(\frac{\sum_{a\in A} x^*_{ja} a}{\sum_{a\in A} x^*_{ja}}\Big)$.
Let $\lambda_{ja} = x^*_{ja} \left/\sum_{a\in A} x^*_{ja}\right.$. 
Then, $\sum_{a\in A}\lambda_{ja} = 1$ for every $j$.
From concavity of the function $p_j(\cdot)$, we have
$$y^*_j\,p_j(b^*_j) = y^*_j\,p_j\Big(\sum_{a\in A}\lambda_{ja} a\Big) \geq 
y^*_j\,\sum_{a\in A}\lambda_{ja} p_j(a) = \sum_{a\in A} x^*_{ja} p_j(a) \geq s_j.$$
\end{proof}

We can further assume that all constraints~($\ref{LP3}'$) are tight i.e., for every 
$j$, we have $y^*_j\,p_j(b^*_j) = s_j$. Indeed, if ($\ref{LP3}'$) is not tight for some $j$, 
then we can decrease $y^*_j$ by letting $y^*_j = s_j/p_j(b_j^*)$. 

\section{Algorithm}
We now describe the approximation algorithm. We first solve the LP relaxation and obtain a solution 
$x^*_{ja}$. Using Claim~\ref{claim:yb}, we convert this solution to the solution $(y^*_j, b^*_j)$
of the simplified LP ($\ref{LP1}'$-$\ref{LP3}'$). We assume that all constraints~($\ref{LP3}'$)
are tight (see above). Then we start the ``rounding'' procedure. 

We schedule jobs iteratively. In every iteration, we schedule the next batch of jobs in the 
interval $[t, t + \Delta t]$ and then advance time from $t$ to $t+\Delta t$. Thus, at the beginning 
of every iteration, we already have a schedule for the time interval $[0, t]$. For every 
job $j$, we keep the remaining size of $j$ in the variable $s^*_j(t)$. Initially, $s^*_j(0)= s_j$.
We also update the LP solution: we maintain variables $y^*_j(t)$ that indicate the time 
required by the remaining portion of job $j$ if $b^*_j$ machines are allotted to it. 
In other words, we maintain the invariant: 
\begin{equation}
\label{eq:invariant-y}
	y^*_j(t) \cdot p_j(b^*_j) = s^*_j(t).
\end{equation}
Initially, $y^*_j(0) = y^*_j$. Hence, for $t=0$, this invariant holds.
 
To schedule the next batch of jobs, we find all unfinished jobs
%% that do not depend on any yet unfinished jobs i.e. those jobs 
that can be scheduled now without 
violating precedence constraints. We call these {\em available} jobs. 
We denote the set of all available
jobs at time $t$ by $\Lambda(t)$. For every available job $j\in \Lambda(t)$  we compute 
$$m^*_j(t) = m\cdot \frac{b^*_j}{\sum_{j\in \Lambda(t)} b^*_j}.$$
We allocate $m_j^*(t)$ machines to job $j$ for the time interval of length  
$$\Delta t = \min_{j\in \Lambda(t)} \frac{s^*_j(t)}{p_j(m_j^*(t))}.$$
Observe that the total number of machines we allocate is $m$.
For all $j\in \Lambda(j)$, we update $s^*_j(t+\Delta t)$ and $y^*_j(t+\Delta t)$:
\begin{eqnarray}
\label{update-s} s^*_j(t+\Delta) & = & s_j^*(t) - p_j(m_j^*(t))\, \Delta t \\
\label{update-y} y^*_j(t+\Delta t) & = & y^*_j(t) - \frac{p_j(m_j^*(t))}{p_j(b^*_j)}\, \Delta t.
\end{eqnarray}
Note that this maintains invariant (\ref{eq:invariant-y}).
We set $t=t+\Delta t$ and proceed to the next iteration.
The algorithm terminates when $s^*_j(t) = 0$ for all $j$.

\section{Analysis}
We now analyze the algorithm. 
First, observe that the algorithm correctly maintains the
remaining sizes $s_j^*(t)$: at time $t$, the remaining size of the job $j$ is indeed $s^*_j(t)$.
Note that all $s^*_{j}(t)$ remain nonnegative (that is how we pick $\Delta t$).
Moreover, at the end of every iteration one of the available jobs, specifically, 
the job $j'$ for which  $\Delta t = s^*_{j'}(t)/p_{j'}(m_{j'}^*(t))$ (again see the definition of $\Delta t$), is completed,
i.e., $s^*_{j'}(t+\Delta t) = 0$.
So the number of iterations of the algorithm is at most $n$, and the running time of the algorithm is 
polynomial in $n$. Also, note that all $y_j^*(t)$ are nonnegative by~(\ref{eq:invariant-y}).

We now need to upper bound the makespan of the schedule produced by the algorithm. We prove the following
standard lemma.

\begin{lemma}\label{lem:C-star}
There exists a chain of jobs $C^*$ such that at every point of time $t$ one and only 
one job from $C^*$ is scheduled by the algorithm.
\end{lemma}
\begin{proof}
Consider the job $j$ that finished last in the schedule generated by the algorithm. We add this job to our chain. This job was not scheduled earlier 
because it depends on some other job $j'$ that finished just before $j$ started. We add $j'$ to our schedule as well. We then pick the job $j'$ depends on, and so on. 
We continue this process until we encounter a job that does not depend on any other job. This job started at time $t=0$. Thus, the jobs in the constructed chain cover
the time line from the beginning to the end of the schedule.
\end{proof}

We now show that for every $t$, the following inequality holds,
\begin{equation}\label{eq:invariant1}
\sum_{j\in C^*} y^*_j(t)  + \frac{1}{m}\,\sum_{j\in J} \, b^*_j y^*_j(t) \leq 2\tilde T -t.
\end{equation}
Note that for $t=0$, the inequality follows from ($\ref{LP1}'$) and ($\ref{LP2}'$). This inequality
implies that the makespan is at most $2\tilde T \leq 2(1+\varepsilon) T$, since 
all $y^*_j(t)$ are nonnegative and thus the left hand side of (\ref{eq:invariant1}) is nonnegative.

\begin{lemma}
Inequality (\ref{eq:invariant1}) holds in the beginning and end of every iteration.
\end{lemma}
\begin{proof}
We assume that (\ref{eq:invariant1}) holds at time $t$ at the beginning of some iteration
and prove that (\ref{eq:invariant1}) holds at time $t+\Delta t$ at the end of this iteration.
In an iteration, the RHS of (\ref{eq:invariant1}) decreases by $\Delta t$. Our goal is to show 
that one of the following happens in any iteration:
\begin{itemize}
\item {\bf Condition (a):} $\sum_{j\in C^*} y^*_j(t)$  (the first term in the LHS of (\ref{eq:invariant1}))
decreases by at least $\Delta t$, or
\item {\bf Condition (b):} $\frac{1}{m}\,\sum_{j\in J} \, b^*_j y^*_j(t)$ (the second term in the LHS
of (\ref{eq:invariant1})) decrease by at least $\Delta t$.
\end{itemize}
Since both the terms in the LHS of (\ref{eq:invariant1}) are non-increasing, the lemma follows.

By Lemma~\ref{lem:C-star}, the algorithm schedules exactly one job in the chain $C^*$ 
in the time interval $[t,t+\Delta t]$. We denote this job by $j'$. By equation (\ref{update-y}), 
we have
$$
y^*_{j'}(t+\Delta t) = y^*_{j'}(t) - \frac{p_{j'}(m_{j'}^*(t))}{p_{j'}(b^*_{j'})} \, \Delta t
= y^*_{j'}(t) - \frac{p_{j'}\Big( \frac{m\, b^*_{j'}}{\sum_{j\in \Lambda(t)} b^*_j}\Big)}{p_{j'}(b^*_{j'})}\, \Delta t.
$$
Denote $\alpha (t) = \sum_{j\in \Lambda(t)} b^*_j$. Rewrite the expression above as follows:
\begin{equation}
\label{eq:y-update-2}
y^*_{j'}(t+\Delta t) 
= y^*_{j'}(t) - \frac{p_{j'}\Big(b^*_{j'}\cdot m/\alpha(t)\Big)}{p_{j'}(b^*_{j'})}\, \Delta t.
\end{equation}

\noindent
{\bf Case 1:} $\alpha(t) \leq m$: %then $y^*_{j'}(t+\Delta t) \leq y^*_{j'}(t) - \Delta t$,
Since $p_{j'}(\cdot)$ is a non-decreasing function and $\alpha(t) \leq m$, we have
$p_{j'}\big(b^*_{j'}\cdot m/\alpha(t)\big)\geq p_{j'}(b^*_{j'})$.
%Hence, if $\alpha(t) \leq m$, then
Using this fact in (\ref{eq:y-update-2}), we have
$$\sum_{j\in C^*} y^*_j(t+\Delta t) \leq \sum_{j\in C^*} y^*_j(t) - \Delta t.$$
Therefore, condition (a) holds in this case.

\noindent
{\bf Case 2:} $\alpha(t) \geq m$: We estimate the second term in the LHS of 
(\ref{eq:invariant1}). Using (\ref{update-y}), we have
\begin{align*}
\frac{1}{m}\sum_{j\in \Lambda(t)} b_j^* y_j^* (t + \Delta t)&=
\frac{1}{m} \sum_{j\in \Lambda(t)} b^*_j \Big( y^*_j(t) - 
\frac{p_j\big(b^*_{j}\,m/\alpha(t)\big)}{p_j(b^*_{j})}\,\Delta t\Big)\\
&=\frac{1}{m}\sum_{j\in \Lambda(t)} b^*_j y^*_j(t) - 
\frac{1}{m}\sum_{j\in \Lambda(t)} b^*_j \,\frac{p_j\big(b^*_{j}\,m/\alpha(t)\big)}{p_j(b^*_{j})}\,\Delta t.
\end{align*}
Since $\alpha(t)\geq m$, we have 
$p_j\big(b^*_{j}\,m/\alpha(t)\big)\geq (m/\alpha(t))\cdot p_j(b^*_{j})$,
since $p_j(\cdot)$ is a concave function with $p_j(0)=0$. Thus,
\begin{align*}
& \frac{1}{m}\sum_{j\in \Lambda(t)} b^*_j y^*_j\cdot (t+\Delta t)
\leq \frac{1}{m}\sum_{j\in \Lambda(t)} b^*_j y^*_j(t) - 
\frac{1}{m}\sum_{j\in \Lambda(t)} b^*_j \,\frac{m p_j(b^*_{j})}{\alpha(t) p_j(b^*_{j})}\,\Delta t\\
&= \frac{1}{m}\sum_{j\in \Lambda(t)} b^*_j y^*_j(t) - 
\sum_{j\in \Lambda(t)}\frac{b_j^*}{\alpha(t)}\,\Delta t
= \frac{1}{m}\sum_{j\in \Lambda(t)} b^*_j y^*_j(t) - \Delta t,
\end{align*}
where the last equation follows from the definition of $\alpha(t)$.
Therefore, condition (b) holds in this case.

Combining the two cases, no matter whether $\alpha(t) \leq m$ or $\alpha(t) \geq m$,
$$
\sum_{j\in C^*} y^*_j(t+\Delta t)  + \frac{1}{m}\,\sum_{j\in J} \, b^*_j y^*_j(t + \Delta t) 
\leq \sum_{j\in C^*} y^*_j(t)  + \frac{1}{m}\,\sum_{j\in J} \, b^*_j y^*_j(t) - \Delta t.
$$
%Note that when we move from time $t$ to time $t+\Delta t$, 
%the right hand side of~(\ref{eq:invariant1}) decreases by $\Delta t$. 
This completes the proof.
\end{proof}

\iffalse
\begin{lemma}
Inequality (\ref{eq:invariant2}) holds in the beginning and end of every iteration.
\end{lemma}
\begin{proof}
Consider an arbitrary iteration and assume that (\ref{eq:invariant2}) holds at the 
beginning of the iteration. We prove that it also holds at the end of the iteration.
For every job $j\in \Lambda(t)$, we have using (\ref{update-y}):
\begin{align*}
& y^*_j(t+\Delta t) p_j(b_j^*) 
= y^*_j(t) p_j(b_j^*) - \frac{p_j(m_j^*(t))}{p_j(b^*_j)} \cdot p_j(b_j^*)\, \Delta t \\
&= y^*_j(t) p_j(b_j^*) - p_j(m_j^*(t))\, \Delta t 
\geq s_j^*(t) - p_j(m_j^*(t))\, \Delta t 
= s^*_j(t+\Delta),
\end{align*}
where the last equation follows from (\ref{update-s}).
\end{proof}
\fi

\section{Analysis for Power Functions}

We now analyze the algorithm for power functions, i.e., 
$p_j(z) = c_j \cdot z^{\gamma}$ for some $\gamma \leq 1$
and constants $c_j > 0$. Let $\delta = 1 - \gamma$.
We now show that for every $t$, the following inequality holds:
\begin{equation}\label{eq:invariant1-power}
\left(\sum_{j\in C^*} y^*_j(t)\right)^{\delta} \cdot \left(\frac{1}{m}\,\sum_{j\in J} \, b^*_j y^*_j(t) \right)^{\gamma} \leq \tilde T-t.
\end{equation}
%and, for every $j$,
%\begin{equation}\label{eq:invariant2}
%y^*_j(t)\,p(b^*_j) \geq s^*_j(t).
%\end{equation}
Note that for $t=0$, the inequality follows from ($\ref{LP1}'$) and ($\ref{LP2}'$).
Inequality~(\ref{eq:invariant1-power}) implies that the makespan is at most $\tilde T$:
all $y^*_j(t)$ are nonnegative, hence the left hand side of inequality~(\ref{eq:invariant1-power}) is also 
nonnegative, consequently $t\leq \tilde T$.
Our main technical tool will be the following fact, which is an easy consequence of 
H\"older's inequality.
\begin{fact}
\label{fact:holder}
	Suppose $X, Y \geq 0$ and $\gamma, \delta \in [0, 1]$ such that $\gamma + \delta = 1$. 
	For any $\Delta X \in [0, X], \Delta Y \in [0, Y]$, define 
	$\Delta \left(X^{\delta}\cdot Y^{\gamma}\right) = X^{\delta}\cdot Y^{\gamma} - (X - \Delta X)^{\delta}\cdot (Y - \Delta Y)^{\gamma}$. Then,
	\begin{equation*}
		\Delta \left(X^{\delta}\cdot Y^{\gamma}\right) \geq (\Delta X)^{\delta}\cdot (\Delta Y)^{\gamma}.
	\end{equation*}
\end{fact}
\begin{proof}
Define vectors 
\iffalse
	\begin{eqnarray*}
		{\bf f} & = & <(X - \Delta X)^{\delta}, (\Delta X)^{\delta}> \\
		{\bf g} & = & <(Y - \Delta Y)^{\gamma}, (\Delta Y)^{\gamma}>. 
	\end{eqnarray*}		
\fi
${\bf f}  =  ((X - \Delta X)^{\delta}, (\Delta X)^{\delta})$ and ${\bf g}  =  ((Y - \Delta Y)^{\gamma}, (\Delta Y)^{\gamma})$.
	Then, by H\"older's inequality, we have
	\begin{equation*}
		\langle {\bf f, g}\rangle \leq \|{\bf  f}\|_{1/\delta} \|{\bf  g}\|_{1/\gamma}
		\Rightarrow
		\quad (X - \Delta X)^{\delta} \cdot (Y - \Delta Y)^{\gamma} + (\Delta X)^{\delta} \cdot (\Delta Y)^{\gamma}
			 \leq X^{\delta} \cdot Y^{\gamma}.
	\end{equation*}
	The lemma follows by rearranging terms.
\end{proof}

Using this fact, we inductively prove that inequality (\ref{eq:invariant1-power})
holds throughout the rounding algorithm.

\begin{lemma}
Inequality (\ref{eq:invariant1-power}) holds at the beginning and end of every iteration.
\end{lemma}
\begin{proof}
We assume that (\ref{eq:invariant1}) holds at time $t$ at the beginning of some iteration
and prove that (\ref{eq:invariant1}) holds at time $t+\Delta t$ at the end of this iteration.
By Lemma~\ref{lem:C-star}, the algorithm schedules exactly one job in the chain $C^*$ 
in the time interval $[t,t+\Delta t]$. We denote this job by $j'$. We have
$$
y^*_{j'}(t+\Delta t) = y^*_{j'}(t) - \frac{p_{j'}(m_{j'}^*(t))}{p_{j'}(b^*_{j'})}\, \Delta t
= y^*_{j'}(t) - \frac{p_{j'}\Big( \frac{m\, b^*_{j'}}{\sum_{j\in \Lambda(t)} b^*_j}\Big)}{p_{j'}(b^*_{j'})}\, \Delta t.
$$
Denote $\alpha (t) = \sum_{j\in \Lambda(t)} b^*_j$. Rewrite the expression above as follows:
$$
y^*_{j'}(t+\Delta t) 
= y^*_{j'}(t) - \frac{p_{j'}\Big(b^*_{j'}\cdot m/\alpha(t)\Big)}{p_{j'}(b^*_{j'})}\, \Delta t
= y^*_{j'}(t) - \left(\frac{m}{\alpha(t)}\right)^{\gamma}\, \Delta t.
$$
%
%
%Note, that if $\alpha(t) \leq m$, then $y^*_{j'}(t+\Delta t) \leq y^*_{j'}(t) - \Delta t$,
%since $p\big(b^*_{j'}\cdot m/\alpha(t)\big)\geq p(b^*_{j'})$ (as $p(\cdot)$ is a non-decreasing function).
%Hence, if $\alpha(t) \leq m$, then
%$$\sum_{j\in C^*} y^*_j(t+\Delta t) \leq \sum_{j\in C^*} y^*_j(t) - \Delta t.$$

We estimate the second term in (\ref{eq:invariant1-power}).
\begin{align*}
& \frac{1}{m}\sum_{j\in \Lambda(t)} b^*_j y^*_j\cdot (t+\Delta t)
= \frac{1}{m} \sum_{j\in \Lambda(t)} b^*_j \Big( y^*_j(t) - 
\frac{p_j\big(b^*_{j}\,m/\alpha(t)\big)}{p_j(b^*_{j})}\,\Delta t\Big)\\
&= \frac{1}{m}\sum_{j\in \Lambda(t)} b^*_j y^*_j(t) - 
\frac{1}{m}\sum_{j\in \Lambda(t)} b^*_j \,\frac{p_j\big(b^*_{j}\,m/\alpha(t)\big)}{p_j(b^*_{j})}\,\Delta t
= \frac{1}{m}\sum_{j\in \Lambda(t)} b^*_j y^*_j(t) - \left(\frac{\alpha(t)}{m}\right)^{\delta}\,\Delta t.
\end{align*}
Using Fact~\ref{fact:holder}, we have
\begin{align*}
& \Delta \bigg(\Big(\sum_{j\in C^*} y^*_j(t)\Big)^{\delta} \cdot \Big(\frac{1}{m}\,\sum_{j\in J} \, b^*_j y^*_j(t) \Big)^{\gamma}\bigg) 
\geq \bigg(\Delta \Big(\sum_{j\in C^*} y^*_j(t)\Big)\bigg)^{\delta} \cdot \bigg(\Delta \Big(\frac{1}{m}\,\sum_{j\in J} \, b^*_j y^*_j(t)\Big) \bigg)^{\gamma} \\ 
&= \quad \bigg(\Big(\frac{m}{\alpha(t)}\Big)^{\gamma}\, \Delta t\bigg)^{\delta} \cdot
\bigg(\Big(\frac{\alpha(t)}{m}\Big)^{\delta}\,\Delta t\bigg)^{\gamma}
= \Delta t.
\end{align*}
Note that when we move from time $t$ to time $t+\Delta t$, the right hand side of~(\ref{eq:invariant1}) decreases by $\Delta t$. This completes the proof.
\end{proof}

\section{Lower Bound for Online Algorithms}
\label{sec:online}

\begin{figure}
\begin{center}
\begin{tikzpicture}
[->,
>=stealth',
shorten >=1pt,auto, node distance=2cm,
thick,
job/.style={circle,fill=red!40!green!10!blue!10,draw},
rjob/.style={circle,fill=red!20,draw}]
  
  \node[job] (u11) {$v_{i1}$};
  \node[rjob] (u12) [below of=u11] {$v_{12}$};
  \node[job] (u13) [below of=u12] {$v_{13}$};
  \node      (u14) [below of=u13] {$\mathbf{\dots}$};
  \node[job] (u15) [below of=u14] {$v_{1l}$};
  
  \node[job] (u21) [right of=u11] {$v_{21}$};
  \node[job] (u22) [below of=u21] {$v_{22}$};
  \node[rjob] (u23) [below of=u22] {$v_{23}$};
  \node      (u24) [below of=u23] {$\mathbf{\dots}$};
  \node[job] (u25) [below of=u24] {$v_{2l}$};
  
  \node[job] (u31) [right of=u21] {$v_{31}$};
  \node[rjob] (u32) [below of=u31] {$v_{32}$};
  \node[job] (u33) [below of=u32] {$v_{33}$};
  \node      (u34) [below of=u33] {$\mathbf{\dots}$};
  \node[job] (u35) [below of=u34] {$v_{3l}$};
  
  \node (u41) [right of=u31] {$\mathbf{\dots}$};
  \node (u42) [below of=u41] {$\mathbf{\dots}$};
  \node (u43) [below of=u42] {$\mathbf{\dots}$};
  \node (u44) [below of=u43] {$\mathbf{\dots}$};
  \node (u45) [below of=u44] {$\mathbf{\dots}$};
  
  \node[job] (u51) [right of=u41] {$v_{k1}$};
  \node[job] (u52) [below of=u51] {$v_{k2}$};
  \node[job] (u53) [below of=u52] {$v_{k3}$};
  \node      (u54) [below of=u53] {$\mathbf{\dots}$};
  \node[job] (u55) [below of=u54] {$v_{kl}$};
  
   \path
   
    (u12) edge node [right] {} (u21)
          edge node [right] {} (u22)
          edge node [right] {} (u23)
			 edge node [right] {} (u25)			 
			 
    (u23) edge node [right] {} (u31)
          edge node [right] {} (u32)
          edge node [right] {} (u33)
			 edge node [right] {} (u35)

    (u32) edge node [right] {} (u41)
          edge node [right] {} (u43)
			 edge node [right] {} (u45);			 			 			 
  \end{tikzpicture}
\end{center}
\caption{The figure shows the execution of an online algorithm. After the last job $v_{i,s}$ in the $i$-th phase is finished (the last job is 
marked with the red color), the adversary presents $l$ new jobs $v_{(i+1)1},\dots v_{(i+1)l}$ that depend only on $v_{is}$.}
\label{fig:online-gap-graph}
\end{figure}
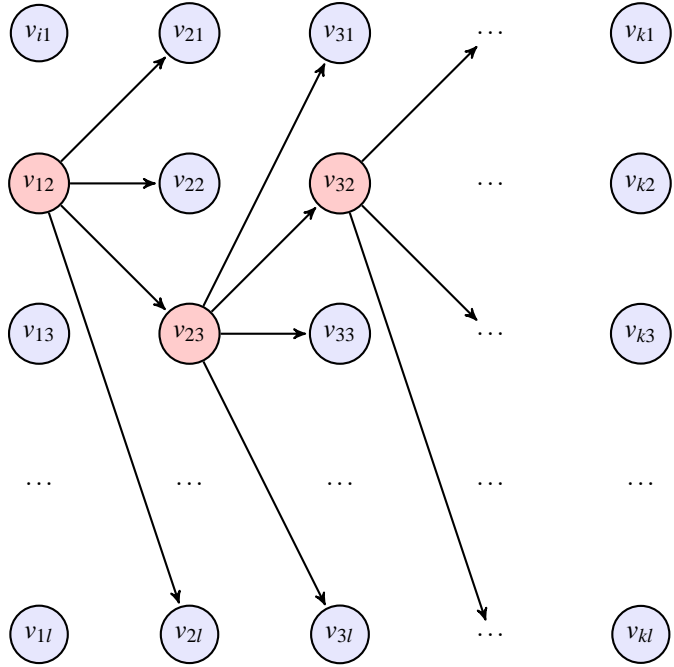

In this section, we show that the \gps~problem has a polynomial lower bound 
in the online setting. Specifically, we show
that even if $p_j(z) = \sqrt{z}$ for all jobs $j$, the competitive ratio of any online algorithm
is at least $\Omega(n^{1/4})$, where $n$ is the number of jobs. Note that in this case 
our offline algorithm gives an almost exact solution ($(1+\varepsilon)$ approximation
for arbitrary $\varepsilon > 0$). In the online model, a job is given to the algorithm only
once all the jobs it depends on are finished. So this result shows that any approximation 
algorithm should use the information about the future schedule and cannot make the decision solely based
on the set of currently available jobs.

We describe the strategy for the adversary. The adversary works in phases. In phase $i$,
she presents $l$ independent jobs $u_{i1},\dots, u_{il}$ to the algorithm. These jobs
depend on the single job in the previous phase that has finished last. That is,
if $u_{ij_i}$ is the job that finished last in the phase $i$, then in
the next phase $(i+1)$, all jobs $u_{(i+1)s}$ depend on this job $u_{ij_i}$
(see Figure~\ref{fig:online-gap-graph}).
We assume that the number of machines is $m=1$.

We lower bound the makespan of the schedule produced 
by the online algorithm. To finish all $l$ jobs given to the algorithm in one phase, we
need to spend time at least $\sqrt{l}$. (The optimal way to allocate machines is to assign $1/l$ machines to each job.)
Thus, the total length of the schedule is at least $k \sqrt{l}$.

Now consider the following solution. Initially, all jobs in the chain $v_{1j_1}, v_{2j_2}, \dots, v_{kj_k}$
are scheduled sequentially (assigning $1$ machine to each job). 
Once all jobs $v_{ij_i}$ are finished, the remaining $kl-k$ jobs are scheduled
in parallel by assigning
$1/(kl-l)$ machines to every job. The length of the schedule equals $k+\sqrt{kl-k}$. 
The lower bound now follows by setting $k=l$ (note that $n=kl$).
%
%Hence, the length of the optimal
%offline schedule is at most $k+\sqrt{kl-k}$.
%The competitive ratio is lower bounded as follows:
%$$\frac{\text{``Online OPT''}}{\text{``Offline OPT''}}
%\geq \frac{k\sqrt{l}}{k+\sqrt{kl - k}} = 
%\frac{\sqrt{kl}}{\sqrt{k}+\sqrt{l - 1}}\geq \frac{\sqrt{kl}}{\sqrt{k}+\sqrt{l}}.$$
%We let $k=l$ and $n=kl$. Then,
%$$\frac{\text{``Online OPT''}}{\text{``Offline OPT''}}\geq \frac{n^{1/4}}{2}.$$
This lower bound can be extended to randomized algorithms using standard techniques,
which we omit for brevity.

We note that this lower bound is almost tight for $p_j(z)=\sqrt{z}$: any algorithm 
that does not idle (i.e., which always allocates all available machines) 
has competitive ratio at most $\sqrt{n}$, since the maximum possible 
rate of processing all jobs is $\sqrt{n}$ (when we process all $n$ jobs in parallel) and
the minimum possible rate is 1 (when we allocate all machines to a single job).

\bibliographystyle{plain}
\bibliography{ref}

\appendix

\section{Polynomial Algorithm for Solving the LP Relaxation}\label{ap:solveLP}

In this section, we describe how one can solve the LP (\ref{LP1}--\ref{LP4}) with accuracy $(1+\varepsilon)$.
We first restrict the set of indices $a$ to $A' = A\cap [\varepsilon/(2m), m]$. If the original LP  has a solution
of cost $\tilde{T}$ then the new LP has a solution of cost $(1+\varepsilon) \tilde T$, because we can move the
mass from variables $x_{ja}$ with $a < a_{min}$ to $x_{a_{min}j}$, where $a_{min} = \min A'$. This change will not
affect constraints~(\ref{LP1}), (\ref{LP3}) and ~(\ref{LP4}):  the left hand sides of (\ref{LP1}) and ~(\ref{LP4})
will not change; the left hand side of (\ref{LP3}) may only increase. The left hand side of  (\ref{LP2}) may also increase, but by no more than 
$a_{min} \tilde T m \leq \varepsilon \tilde T$. So the constraint is valid if we replace $\tilde T$ with $(1+\varepsilon) \tilde T$.

Now, the only problem is that the LP has exponentially many constraints of the form:
$$\sum_{j\in C}\sum_{a\in A} x_{ja} \leq \tilde T \;\;\text{ for every chain } C$$
These constraints can be rewritten using polynomially many constraints by introducing axillary variables.
Alternatively, we can use the ellipsoid method to solve the LP. The separation oracle for 
these constraints needs to check that every chain has length at most $\tilde{T}$, or in other words, 
that the maximum chain has length at most $\tilde{T}$. The maximum chain can be found in polynomial time
(see e.g., the book of Schrijver~\cite{schrijver}, Section~14.5 for more details). 

\end{document}